\documentclass[12pt]{article}
\input epsf
\begin{document}
\thispagestyle{empty}
\begin{center}

{\Large\bf Do we understand the single-spin asymmetry for $\pi^0$ inclusive 
production in $pp$ collisions?}\\

\vskip1.4cm
Claude Bourrely and Jacques Soffer  
\vskip0.3cm
Centre de Physique Th\'eorique\footnote{Unit\'e propre de  Recherche 
7061}, CNRS-Luminy, \\Case 907, F-13288 Marseille Cedex 9 - France
\vskip 2cm
{\bf Abstract}\end{center}
The cross section data for $\pi^0$ inclusive production in $pp$ collisions 
is considered in a rather broad kinematic region in energy $\sqrt{s}$, 
Feynman variable $x_F$ and transverse momentum $p_T$. The analysis of these 
data is done in the perturbative QCD framework at the next-to-leading 
order. We find that they cannot be correctly described in the entire kinematic
domain and this leads us to conclude that the single-spin asymmetry, $A_N$ 
for this process, observed several years ago at FNAL by the experiment E704 
and the recent result obtained at BNL-RHIC by STAR, are two different 
phenomena. 
\vskip 2cm 
\noindent PACS numbers: 12.38.Bx, 13.85.Ni, 13.88.+e
\vskip 2cm
\noindent CPT-2003/P.4581

\begin{figure}
\begin{center}
\leavevmode {\epsfysize= 17.cm \epsffile{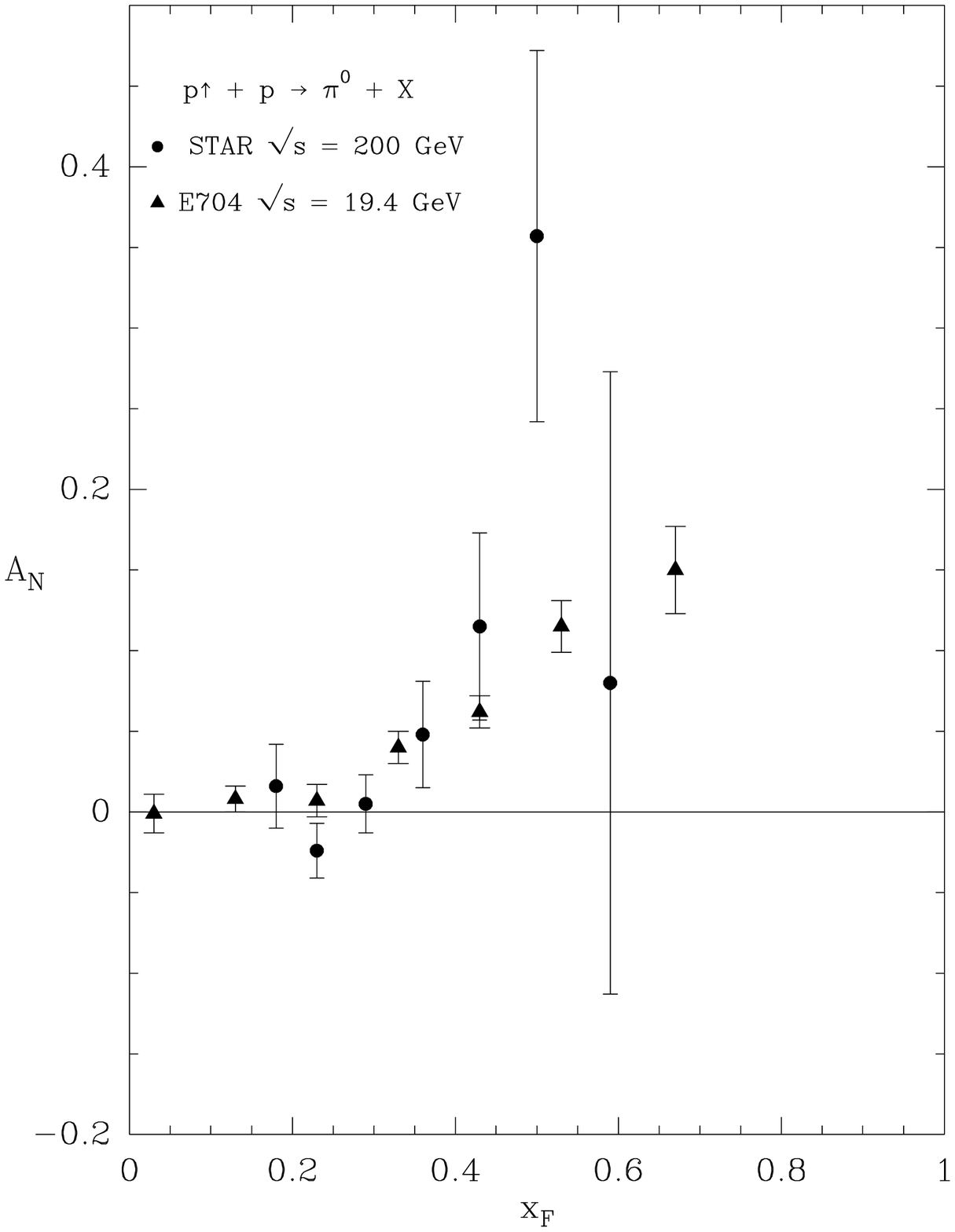}}
\end{center}
\caption[*]{\baselineskip 1pt
The single-spin asymmetry $A_N$ as a function of $x_F$, at two different 
energies. The data are from Refs. \cite{E704a,STAR}.
}\label{fig1}
\end{figure}

\newpage
High-energy single-spin asymmetries (SSA), usually denoted by $A_N$, in pion 
inclusive production $pp^{\uparrow} \to \pi X$, have been observed for 
the first time more than ten years ago, by the Fermilab E704 experiment 
\cite{E704,E704a} at $p_{lab}=200 \mbox {GeV/c}$ 
($\sqrt{s} =19.4 \mbox {GeV}$), in the beam fragmentation region.
For clarity, we recall that $A_N$ is defined as
\begin{equation}
A_N= \frac {d\sigma(\uparrow) - d\sigma(\downarrow)}
{d\sigma(\uparrow) + d\sigma^(\downarrow)}~,
\label{A_N}
\end{equation}
where $d\sigma(\uparrow)$,$(d\sigma(\downarrow))$  is the $\pi$
production cross section with one proton beam
transversely polarized in the {\it up} ({\it down}) direction, with
respect to the normal to the scattering plane.  
A striking dependence of $A_N$ in Feynman variable $x_F$ was observed, 
as shown in Fig. \ref{fig1},for the neutral pion case, $p p^{\uparrow} 
\to \pi^0 X$. Very recently, the STAR Collaboration \cite{STAR} at
BNL-RHIC has released new data for the same SSA at 
$\sqrt{s} = 200 \mbox {GeV}$, also depicted in Fig. \ref{fig1}. 
It is remarkable that these two sets of data cover approximately the same 
$x_F$ range and, assuming that they have the same dynamical origin, 
it might be tempting to conclude that this SSA is energy independent, 
to a first approximation.

On the theoretical side, according to naive parton model arguments one expects
$A_N = 0$, but several possible mechanisms have been proposed recently to 
generate a non-zero $A_N$. They are based on the introduction of a transverse 
momentum ($k_T$) dependence of either the distribution functions, for the 
Sivers effect \cite{DS} or of the fragmentation function, for the 
Collins effect \cite{JCC}. These leading-twist QCD mechanisms have been used 
for a phenomenological study of this SSA \cite{AM} and higher-twist effects
have been also considered \cite{QS,KK,EKT}. However a simultaneous description
of $A_N$ and the unpolarized cross section has been ignored in all these
works, a relevant point which has been already emphasized \cite{JS}
and which will be crucial in the remaining of this paper.

Our starting point is the analysis of the cross section data in a broad 
kinematic region in the framework of perturbative QCD (pQCD), at 
next-to-leading order (NLO). It is a common belief that it should be 
best working in the central region, namely, for a $\pi^0$ produced near 
90$^o$, in the center-of-mass system, although some authors have 
argued that the phenomenological inclusion of intrinsic
$k_T$ effects leads to a better agreement between theory and data \cite{E706}. 
The inclusive invariant cross section for the reaction $p p \to \pi^0 X$ reads

\begin{eqnarray}
E_{\pi} d\sigma / d^3p_{\pi}
= \sum \limits_{abc} \int dx_a dx_b f_{a/p}(x_a,Q^2, \mu_F) \times  
\nonumber \\ 
f_{b/p}(x_b,Q^2, \mu_F){ D_{\pi^0/c}(z_c,Q^2, \mu'_F) 
 \over \pi z_c}d\hat {\sigma} /d \hat{t}(ab \to cX)~,
\label{Dsig}
\end{eqnarray}
where the sum is over all the contributing partonic channels $a b \to c X$ 
and $d \hat {\sigma}/ d \hat{t}$ is the associated partonic cross 
section. In our calculations the parton distribution functions (PDF) 
$f_{a/p},f_{b/p}$ are the BBS set constructed in Ref. \cite{bou02} 
and $D_{\pi^0/c}$ is the pion fragmentation function BKK
of Ref. \cite{BKK} \footnote{We have checked that the KKP fragmentation 
functions of Ref. \cite{KKP} give close numerical results}.
$\mu_F$ and $\mu'_F$ denote two factorization scales and there is also 
a renormalization scale $\mu_R$, associated to the running strong coupling 
constant $\alpha_s$. Our calculations are done up to the NLO corrections, 
using the numerical code INCNLL of Ref. \cite{AVER}, with  
$\mu_F = \mu'_F = \mu_R = \mu $.

\begin{figure}
\begin{center}
\leavevmode {\epsfysize= 17.cm \epsffile{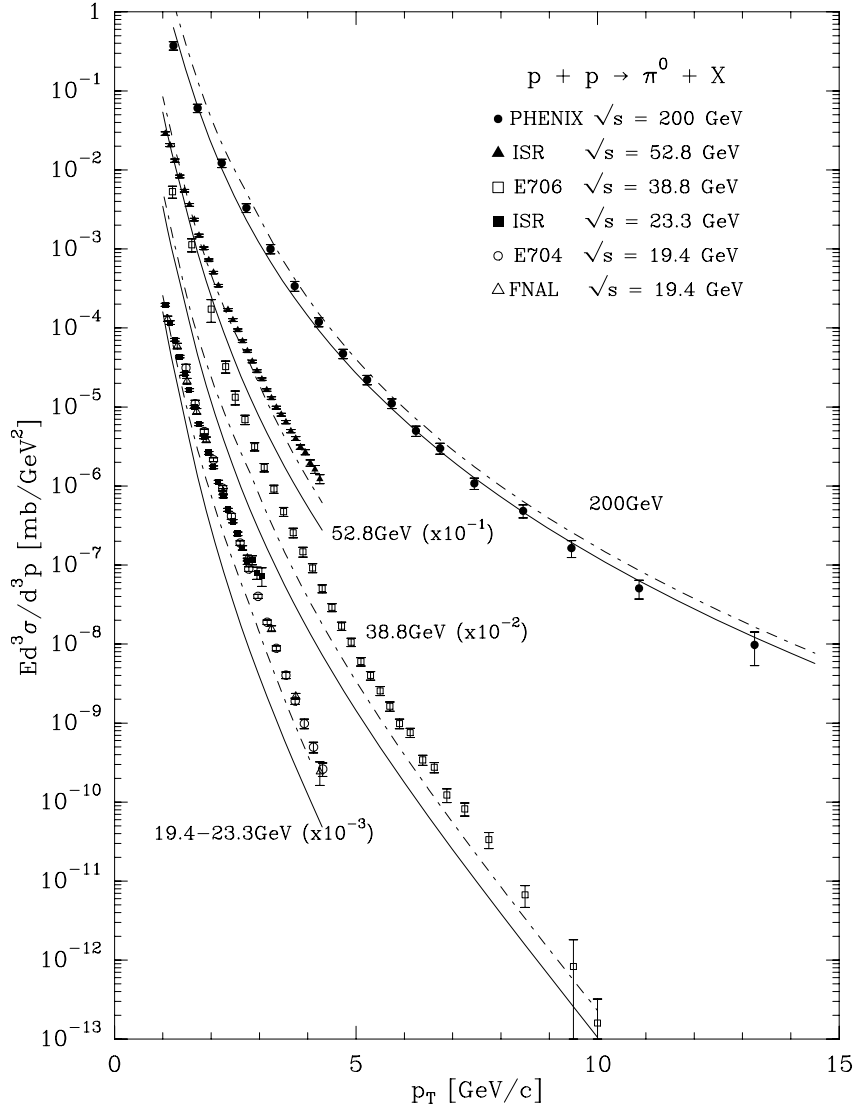}}
\end{center}
\caption[*]{\baselineskip 1pt
$Ed^3\sigma/d^3p$ at 90$^o$ and various energies, as a function of $p_T$. 
Data are from Refs. \cite{E706,E704b,PHE,ISR,BCL} and the 
curves are the corresponding NLO pQCD calculations with
$\mu = p_T$ (solid lines) and $\mu = p_T/2$ (dotted- dashed lines).
}\label{fig2}
\end{figure}

First we look at the cross section data at 90$^o$ for various energies, 
including data from ISR, Fermilab and BNL-RHIC, between 
$\sqrt{s} = 19.4 \mbox {GeV}$ and $200 \mbox {GeV}$, as a function of $p_T$ 
and the comparison with our calculations are displayed in Fig. \ref{fig2}.
As it is well known, the agreement between theory and experiment depends on 
the choice of $\mu$ and this was extensively discussed, for example in 
Ref. \cite {Aur00}. For comparison we show the results for two
scales $\mu = p_T$  and $\mu = p_T/2$ and we see that systematically the 
cross section for $\mu = p_T$ lies under that for $\mu = p_T/2$. 
In our case the best agreement occurs with $\mu = p_T$ for 200GeV and 
$\mu = p_T/2$ for 52.8GeV and below. We get a beautiful description of the 
PHENIX data but we have some disagreement at lower energies; 
we underestimate the data by a factor two on average and more for E706, 
in particular, at low $p_T$ values. It is important to recall that the E704 
Collaboration has also measured $A_N$ in this kinematic region 
$\theta = 90^o$ and found $A_N = 0$ \cite{E704b}, up to 
$p_T = 4.5 \mbox {GeV}/c$. The same trend has been observed
recently by the PROZA-M experiment at $p_{lab} = 70 \mbox {GeV}/c$ 
\cite {PROZA} and all these results are consistent with an earlier theoretical
bound \cite{BS77}.

It is also crucial to note that in Fig. \ref{fig1}, 
whereas the STAR data points are at a fixed angle $\theta = 2.6^o$, 
this is not the case for the E704 data, for which 
$9^o < \theta < 67^o$, due to the simple relation 
$x_F = 2p_T/\sqrt{s} \, \mbox{tan}\theta$ and the fact that $<p_T>$ lies
between 0.7 and 1 GeV/c. In this
case, one can check that $A_N = 0$  for the three lowest $x_F$ points
which correspond to $\theta > 15^o$, whereas $A_N$ is not zero only for the 
four highest $x_F$ points which correspond to $\theta < 15^o$. 

\begin{figure}
\begin{center}
\leavevmode {\epsfysize= 17.cm \epsffile{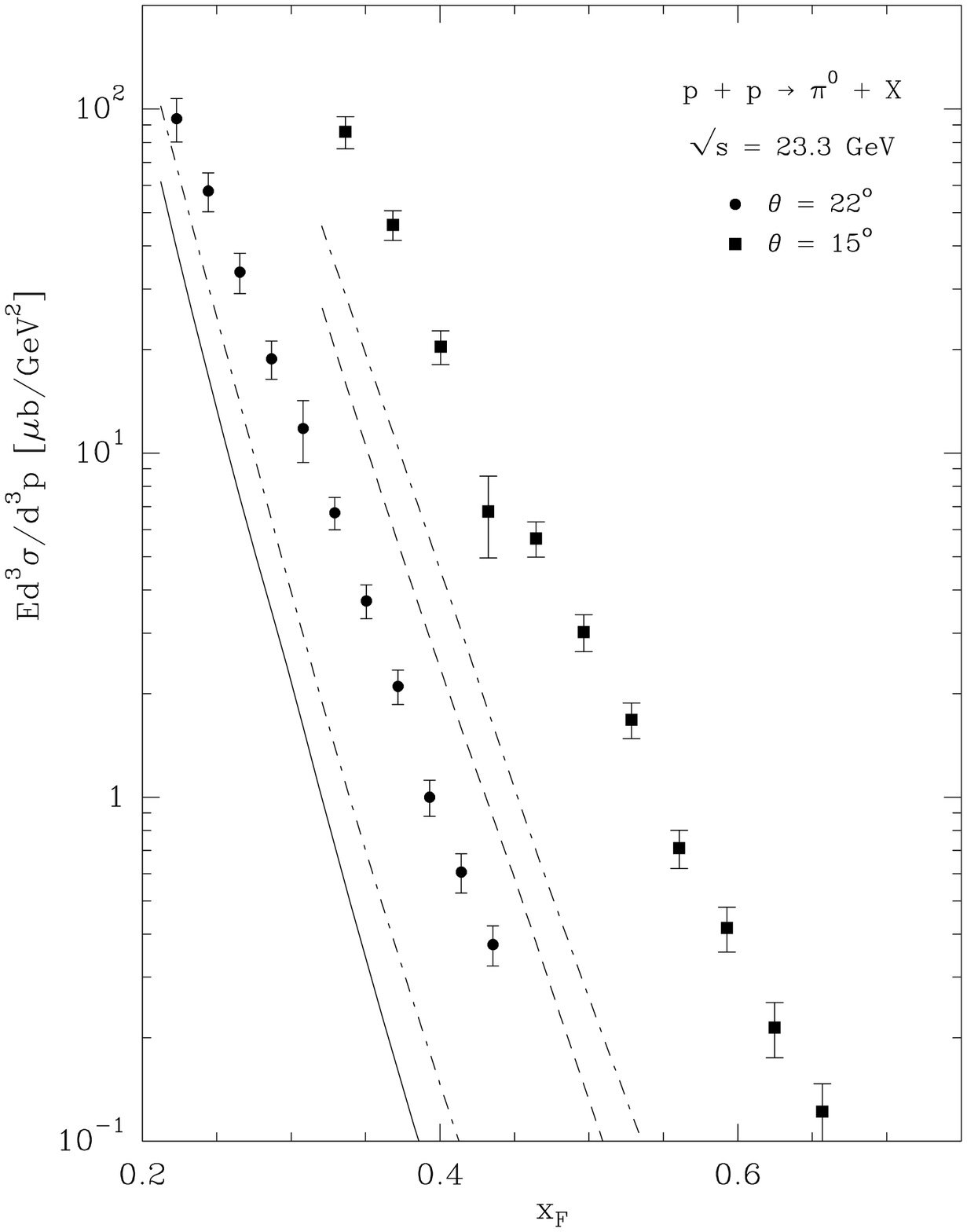}}
\end{center}
\caption[*]{\baselineskip 1pt
$Ed^3\sigma/d^3p$ at $\sqrt{s}=23.3 \mbox{GeV}$, as a function of $x_F$ for two
different scattering angles. The data are from Ref. \cite{ISR} and the curves,
solid $\theta$= 22$^o$ and dashed $\theta$= 15$^o$, are the corresponding 
NLO pQCD calculations with $\mu = p_T$. The dotted-dashed curves are for 
$\mu = p_T/2$.
}\label{fig3}
\end{figure}

\begin{figure}
\begin{center}
\leavevmode {\epsfysize= 17.cm \epsffile{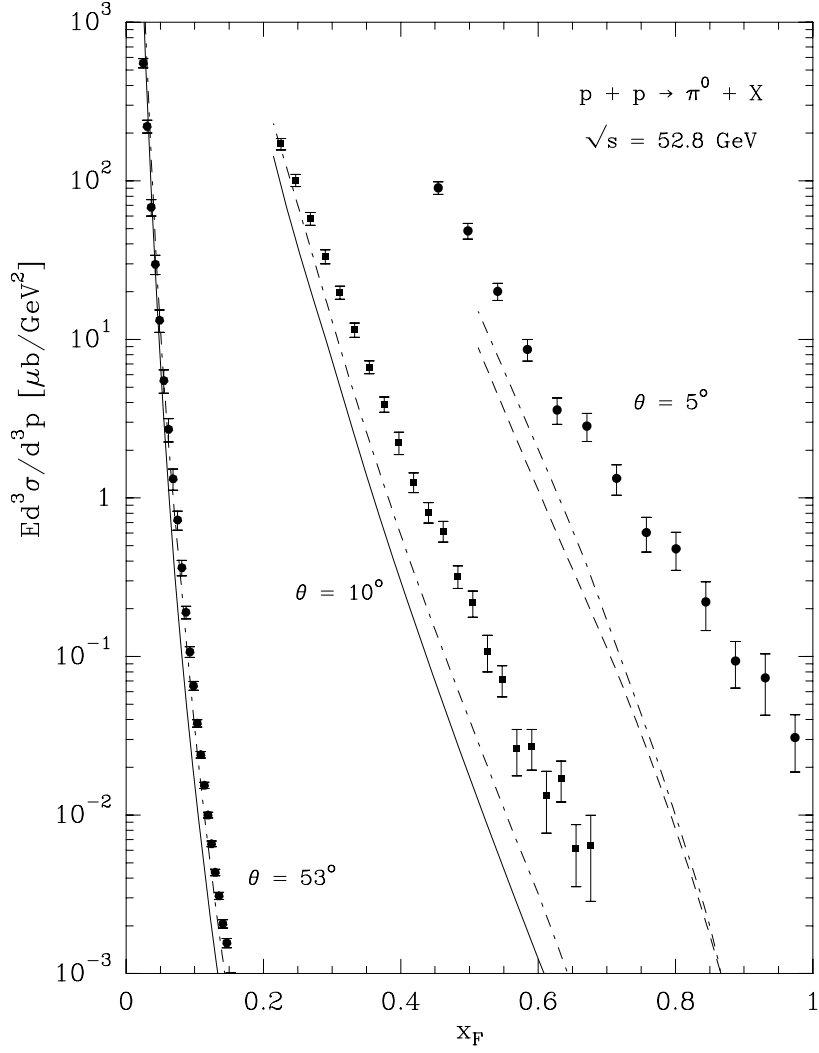}}
\end{center}
\caption[*]{\baselineskip 1pt
$Ed^3\sigma/d^3p$ at $\sqrt{s}=52.8 \mbox{GeV}$, as a function of $x_F$ for 
three different scattering angles. The data are from 
Ref. \cite{ISR} and the curves are the corresponding NLO pQCD calculations 
with $\mu = p_T$. The dotted-dashed curves are for $\mu = p_T/2$.
}\label{fig4}
\end{figure}

This observation led us to examine the unpolarized cross sections away from 
$\theta = 90^o$, as a function of $x_F$, which combines the effects of $p_T$ 
and $\theta$. We have considered the ISR data \cite{ISR} which have
the largest angle coverage, they are shown in Figs. \ref{fig3} and \ref{fig4},
with our numerical calculations. It is striking to observe that the NLO pQCD 
results give a smaller and smaller fraction of the measured
cross section, when going to smaller and smaller scattering angles. 
At $\sqrt{s} = 23.3 \mbox{GeV}$, which is very close to the E704 energy, for 
$\theta = 15^o$ the ratio Data/Theory can be one order of magnitude, even for 
$\mu = p_T/2$, which means that another mechanism is at work and dominates, 
very likely related to soft processes of some specific nature. 
At $\sqrt{s} = 52.8 \mbox{GeV}$, we get a perfect description of 
the $\theta = 53^o$ data, for $\mu = p_T/2$, but again the relative size of 
the soft contributions increases in the forward direction, where the 
disagreement between data and pQCD gets larger. The failure of the NLO pQCD
to reproduce the ISR small angle data is certainly not due to the
breakdown of the convergence of the pQCD series in this region. Indeed the
" K factor ", defined as usual as $d\sigma_{NLO}/d\sigma_{LO}$, where LO
stands for leading-order, is larger
for small angles. We have also observed that there is a considerable
reduction of the scale dependence, going from LO to NLO which shows an
improvement in the perturbative stability. Therefore next-to-next-to-leading
or higher pQCD corrections are not expected to account for a factor ten or so
in the ratio Data/Theory.

\begin{figure}
\begin{center}
\leavevmode {\epsfysize= 17.cm \epsffile{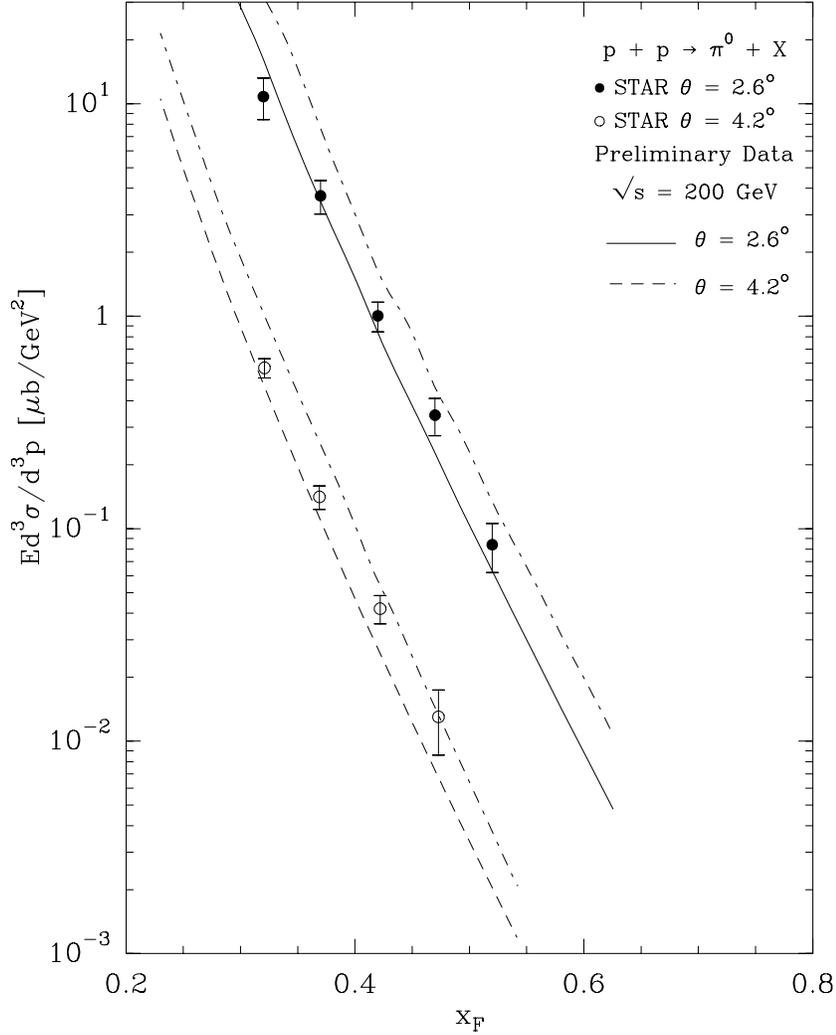}}
\end{center}
\caption[*]{\baselineskip 1pt
$Ed^3\sigma/d^3p$ at $\sqrt{s}=200 \mbox{GeV}$, as a function of $x_F$. The 
solid and dashed curves are the NLO pQCD calculations, with $\mu = p_T$, 
at two different angles and the data points are from
Ref. \cite{STAR}. The dotted-dashed curves are for $\mu = p_T/2$.
}\label{fig5}
\end{figure}

Finally, we show in Fig. \ref{fig5}, the NLO pQCD calculations near 
the forward direction at $\sqrt{s} = 200 \mbox {GeV}$ 
and one notices that at this energy, the $\theta = 2.6^o$ STAR data is very 
well reproduced by the solid curve ($\mu = p_T$). For the sake of completeness
we also give our prediction (dashed curve) at 
$\theta = 4.2^o$, compared to the preliminary data released recently by
the STAR Collaboration \cite{STAR}. 
Now let us go back to the SSA measured by E704 and STAR shown in Fig. 
\ref{fig1}. In both cases, the data lie near the forward direction, 
but in the case of E704 it cannot be attributed to a pure pQCD mechanism, 
because the measured unpolarized cross section can be one
order of magnitude higher than the pQCD result (see Fig. \ref{fig3}). 
On the other hand, it is very likely to be so in the case of 
STAR which has measured the cross section in full agreement with pQCD. 
Therefore this analysis strongly suggests that the two sets of SSA data, 
we have considered here, are the manifestation of two different 
phenomena, in contrast with the implications from Refs. \cite{AM,QS,YK}.

To conclude, if pQCD does not generate a SSA in the $\theta = 90^o$ region, 
as shown by E704, we dare to predict that PHENIX should also find 
$A_N = 0$, a result, indeed perfectly consistent with the recent rigorous 
positivity bound \cite{JSb}, which implies $|A_N| \leq 1/2$. It will be very 
interesting to know what the data analysis will give, when completed in 
the near future \footnote{ This has been confirmed by very recent
preliminary data \cite{Aid}.}.\\ 

{\it Note added :}  After this paper was completed, L. Bland and A. Ogawa have
compared PYTHIA calculations of the invariant cross section with STAR and ISR
data and they confirm our conclusions.

\vskip 0.3cm \noindent {\bf Acknowledgments:} 
We thank W. Vogelsang for checking our numerical results and a very useful 
discussion. We are also grateful to M. Begel, G. Bunce, L. Bland and N. Saito
 for interesting comments and suggestions.

\newpage
\end{document}